\documentclass[12pt]{article}
\def\rr{{{\bf r}}}
\def\rrp{{{\bf r^\prime}}}
\def\nup{{{n_\uparrow}}}
\def\ndo{{{n_\downarrow}}}

\begin{document}

\title{ Study of interaction of high-power Ar$^+$ laser beam with Ag$^+$-doped
glass}

\author{A. Nahal$^\dagger$, H. R. M. Khalesifard$^\dagger$, and M.
Payami$^{\dagger,\ddagger}$}

\maketitle

\begin{center}
{$\dagger$-\it Department of Physics, Institute for Advanced
Studies in Basic Sciences (IASBS),\\ P.O.Box 45195-159,
Zanjan-Iran}
\end{center}
\begin{center}
{$\ddagger$-\it Center for Theoretical Physics and Mathematics,
Atomic Energy Organization of Iran,\\ P.~O.~Box 11365-8486,
Tehran, Iran}
\end{center}

\begin{abstract}
In this work, we have investigated the interaction of a
high-power Ar$^+$ laser beam, in a continuous multi-line regime,
with Ag$^+$-doped glass samples. The samples were subjected to the
irradiation after the ion-exchange step. As a result of the
irradiation, a peak appears in the absorption spectrum; its
evolution depends on both the exposure time and the laser beam
power. Interaction of the beam with the clusters causes them break
into smaller ones which has been confirmed by optical absorption
spectroscopy. Depending on the increment steps of the laser-beam
power, different results are obtained. It is found that, in
addition to the fragmentations of the nano-clusters, clusters of
micro-meter size ($d\sim 3\mu m$) are formed in the sample, if
the laser power is increased in a fast regime. The
 fragmentation processes have been explained in the
framework of the density-functional theory with stabilized
jellium model.

\end{abstract}
\newpage

\section{Introduction}
\label{sec1}

Nonlinear optical materials, such as glasses doped with metallic
clusters, are promissing ones for optical switching technology
\cite{Vogel89}. These materials, because of their high
transparency; ease of fabrication; durability and
thermomechanical stability, are essential component parts of
functional devices for optical communications, sensing and
computing\cite{Kobayashi,Friberg,Stegeman}. Metal nano-cluster
composites are particularly interesting because, depending on the
choice of the metal and the dielectric host, they exhibit
different pico-second relaxation times, range of chemical
reactivities, and photo-sensitivities\cite{Gonella}

Several methods have been reported on producing and manipulating
of metal-cluster doped-glasses: Low-mass ion
irradiation\cite{Caccavale,Battaglin96,Battaglin,Gonellanucl},
annealinig in hydrogen atmosphere after the ion-exchange (IE)
process\cite{Marchi}, and induced modifications of nano-clusters
by pulsed laser beam\cite{Gonella}.

In this paper, we report the results of interaction of a CW
high-power Ar$^+$ laser beam with Ag$^+$-doped glass samples, in
two regimes of ``slow'' (SR) and ``fast'' (FR). In the SR, the
power of the pump beam has been increased from $0.1\;mW$ to
$8\;W$ in a period of $\sim90$ minutes ( $0.5\;W$ increment per 5
minutes ). In the FR, the power is increased up to $8\;W$ in 1
minute. The irradiation of the samples has been performed after
the IE step.

The outcome of the SR results show that a 410 nm peak is appeared
in the absorption spectrum and the peak in the UV region of the
spectrum has been blue-shifted relative to that of non-irradiated
samples. The appearance of the 410 nm band peak implies the
generation of neutral Ag clusters\cite{Kreibig}, and the
blue-shift of the peak in the UV region implies the generation of
smaller clusters\cite{Kresin}. Both these features can be
explained on the basis of possible fragmentation processes of the
ionized nano-clusters. The evolution of these peaks are found to
depend on the exposure ($H=I.t$, where $I$ is the intensity and
$t$ is the exposure time).

In the FR, also the 410 nm peak is appeared but, in this case,
the UV peak is blue-shifted relative to that in the SR results.
In addition, in the FR, we have observed the generation of
micro-clusters. These experimental facts are discussed in detail
in the next section.

The organization of this paper is as follows. In section \ref{sec2} we explain
the details of the experiments and discuss the results. Section \ref{sec3} is
devoted to the density-functional calculations of the possible fragmentations
of singly-ionized and doubly-ionized
silver clusters, and finally, we conclude our work in section \ref{sec4}.

\section{Experiments and discussions}
\label{sec2}

In our experiments, the samples are prepared using the well-known
ion-exchange (IE) method. In the IE process, some of the Ag$^+$
ions in the molten salt and the Na$^+$ ions in the glass matrix
are exchanged by diffusion through the interfaces at both
sides\cite{Najafi}. The diffusion depth and the concentration of
the Ag$^+$ ions in the glass matrix depend on the temperature of
the molten salt, the IE duration, and the Ag$^+$ concentration in
the molten salt. In this experiment, we insert a glass slide
($20\times 30\times 1.8$ mm)into the molten NaNO$_3$/AgNO$_3$
mixed-salt ($98\% / 2\%$) of temperature 320 $^\circ C$ for 4
hours (Fig. \ref{fig1}). The glass components of our samples are
specified in Table \ref{tab1}. The setup used to study the
interaction of Ar$^+$ laser (pump beam) with the prepared samples
is sketched in Fig. \ref{fig2}. In this setup, the samples are
subjected to the pump beam at the angle of incidence close to the
normal. We have performed the experiments in two regimes of
irradiation: The SR and the FR. Depending on which regime is
chosen, different outcomes result in.

One of the common features observed in both regimes, during the
irradiation process, is the formation of a micro-lens at the
incidence point due to thermal effects induced by the pump beam.
Formation of the micro-lens is implied by the appearance of a
circular interference pattern on the screen (Fig. \ref{fig2} top
left). This pattern is generated due to interference of the
reflected beams from the front and back sides of the sample,
which have obtained different radii of curvatures (In the SR, the
formation of this micro-lens is initiated beyond a threshold of
$P_t\sim 4$ W). For a fixed intensity of the pump beam, the radii
of these pattern rings increase with the exposure time up to a
limiting value for which a local thermal equilibrium has been
achieved. The other common feature, observed in both regimes, is
the appearance of a 410 nm band peak and weakening of the peak at
the UV region of the absorption spectra (Fig.\ref{fig3}).In this
work, all spectra have been measured using a {\it Varian Cary-5}
UV-Vis-NIR spectrophotometer. The dimension-less quantity $D$ in
the absorption spectra refers to the optical density
\begin{equation}\label{ddef}
  D=-\log(\frac{I}{I_0}),
\end{equation}

where, $I$ is the intensity of the probe beam and $I_0$ is the
intensity of the reference beam in the spectrophotometer. The 410
nm peak is resulted from the excitation of the surface plasmons
of neutral silver clusters\cite{Kreibig}.  The neutral silver
clusters can be created through different possible processes. In
a fragmentation process, an ionized cluster may decay to smaller
parts, one of which could be a neutral cluster. These
fragmentation processes are discussed in detail in section
\ref{sec3}. Another process could be the capture of the electron,
detached from another cluster.

However, there are remarkable differences in the spectra of the
two regimes: In the FR (curve 1 of Fig. \ref{fig4}), the 410 nm
peak is narrower and higher than that of the SR (curve 2 of
Fig.\ref{fig4}). In addition, a blue-shift is modulated on the UV
peak of the FR spectra compared to that of the SR. The blue-shift
in the spectra of the FR implies the production of smaller
clusters due to fragmentation\cite{Kresin} The difference in the
shapes of the 410 nm peaks in the two regimes originates from the
different size distributions of the clusters in the irradiated
samples. In general, a sharper peak implies a sharper size
distribution of clusters.

\subsection{Results under the Slow Regime}
Under the SR condition, we notice the creation of neutral Ag
clusters, which is implied by the appearance of a 410 nm band
peak in the absorption spectra (Curve 2, Fig. \ref{fig4}).

To analyze the spectra from different parts of the interaction
region, we have divided the circular-shape interaction region
into smaller sub-regions, along the diameter, separated by $0.2\;
mm$ and have measured their respective spectra. The results are
shown in Fig. \ref{fig5}. Labels 1, 2, $\ldots$, 12 in Fig.
\ref{fig5} correspond to the sub-regions from the outer parts
towards the center of the interaction region, respectively. As is
seen, with moving towards the center, the height of the 410 nm
peak increases and at the same time, the height of the UV-region
peak decreases. This decrease is accompanied by a blue-shift.
These behaviors are explained as follows. It is well-known that
the intensity profile across the laser beam is Gaussian, i.e., the
central part of the interaction region receives more photons than
the outer parts. Knowing this fact, and considering the curves 1,
2, $\ldots$, 12 of Fig. \ref{fig5} one can argue that, the more
photons received by the sample glass, the more fragmentations of
the clusters are induced. As a result of these fragmentations,
the neutral Ag clusters are produced. We will show in section
\ref{sec3} that for large ionized clusters the fragmentations in
which one of the products is neutral, are most favored
(evaporation processes).

In Fig. \ref{fig6}, we have plotted the height of the 410 nm
absorption peak as a function of the radial distance in the
interaction region. The heights are normalized to the maximum
height at the center. The center of the interaction region is
located at point 2.4 mm (The size of the interaction region is
taken to be 4.8 mm). This plot re-confirms the Gaussian intensity
distribution.

The relative blue-shifts of the peaks at the UV region of the
absorption spectra in different sub-regions (Fig. \ref{fig5}),
imply that the sizes of the clusters decrease\cite{Kresin} as we
move towards the central sub-region (the UV peak corresponds to
the volume plasmon excitations of Ag clusters [See Ref.
\ref{Kreibig}, page 35]). At the same time, one sees that the
height of the UV peak decreases by moving towards the central
sub-region. This weakening is due to the decrease in the
concentration of clusters as a result of interaction with the
pump beam. In fact, the pump beam causes the clusters fragment
and scatter in such a way that the conservation of momentum is
satisfied.

Some of the mechanisms underlying the production of the neutral
clusters, responsible for the 410 nm peak, could be the following
chain fragmentation:
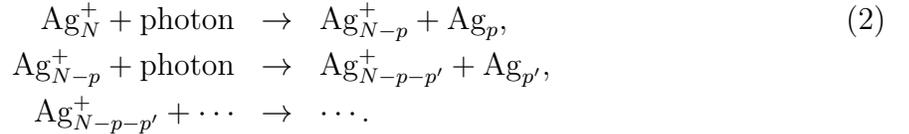
\begin{eqnarray}\label{eeqq1}
{\rm Ag}_N^+ +{\rm photon}&\rightarrow &{\rm Ag}_{N-p}^+ + {\rm
Ag}_p,\\ \nonumber {\rm Ag}_{N-p}^+ +{\rm photon}&\rightarrow &
{\rm Ag}_{N-p-p^\prime}^+ + {\rm Ag}_{p^\prime},\\ \nonumber {\rm
Ag}_{N-p-p^\prime}^+ + \cdots &\rightarrow& \cdots.
\end{eqnarray}
The fragmentation of large multiply ionized clusters (if any) are
less favored because, they require photons with higher energies.
The theoretical calculations related to the fragmentations of
silver clusters has been discussed in section \ref{sec3}.

Another mechanism for production of the neutral clusters is
recombination of the singly ionized clusters with the electrons
detached from other possible neighboring sources (glass matrix):
\begin{equation}\label{eeqq2}
{\rm Ag}_M^+ +{\rm e}^-\rightarrow {\rm Ag}_M.
\end{equation}
Heating the non-irradiated ion-exchanged samples, under the normal
atmosphere conditions for temperature beyond the ion-exchange
temperature, leads to the appearance of 410 nm peak in the
absorption spectra and a red-shift in the  UV peak of the
absorption spectra (Fig. \ref{fig7}). This experimental
observation justifies the mechanism of Eq. (\ref{eeqq2}).

\subsection{Results under the fast regime}

Under the FR condition, as under the SR, the creation of the 410
nm peak is observed. Comparing the absorption spectra of the
central parts of the interaction regions in the two regimes,
reveal that in the FR, this peak is narrower and higher; and in
addition, the UV peak is blue-shifted as well as decreased
relative to that in the SR results (Curve 1, Fig. \ref{fig4}). As
mentioned before, a sharper peak corresponds to a sharper size
distribution of clusters, and the blue-shift of the UV peak is
related to the production of smaller clusters. To understand the
detailed mechanisms responsible for these differences between the
SR and the FR, more experimental investigations are needed.

In addition to the above-mentioned differences, under the FR, we
have observed the generation of micrometer clusters. This
experimental fact is discussed as follows.

After the irradiation under the FR, the appearance of large
($\sim\mu m$) neutral Ag clusters has been observed at both sides
of the sample. The distributions of the $\mu$m-clusters at the
two sides have significant differences. The observations have
been done by an optical microscope using bright-field photography
method.

On the front side, the neutral $\mu$m-clusters are resided in a
ring-shape area as islands. Each of these islands is formed, at a
nucleation center, by absorbing its surrounding nano-clusters
[Fig. \ref{fig8}(a)-(c)]. The nucleation centers mostly lie in
relatively cold regions in the glass matrix which corresponds to
the lower-intensity parts of the incidence area (the laser beam
has a gaussian profile of intensity). However, because of the
inhomogenity of the glass matrix, some parts in the inner region
have also higher viscousities\cite{Vogel89,Mazurin} which serve
as nucleation centers. Due to the radial gradient in the
laser-beam intensity, the radiation pressure varies in the radial
direction which, in turn, cause the relatively larger clusters
($d\gg \lambda/20$)in the central-part be pushed away from the
center\cite{Sasaki,Svoboda,Zemanek}. The radial velocities of the
clusters decrease with distance from the center. Thus, they can
aggregate mostly in the ring-shape region. On the back side of
the sample, on the other hand, the $\mu$m-clusters are seen to be
distributed more or less uniformly in a disc-shape area, centered
at the beam center [Fig. \ref{fig8}(d), (e)]. This difference in
the distributions at the front and back sides can easily be
explained by comparing the beam intensities at these two sides.
On measuring the transmittance of the beam for each line of the
laser beam, we conclude that during the irradiation, the
intensity at the back side is much less than the front side
(Table \ref{tab2}), so that the mobility of the clusters is much
less than that of the front side. This cause the back-side
clusters not to be pushed away from the center but, just
aggregate at those nucleation centers which correspond to the
more viscous parts of the glass matrix\cite{Vogel85}. To our
knowledge, the formation of the $\mu$m-sized silver clusters in
the glass matrix under the action of a CW laser beam has not been
reported.

\section{Theoretical study of the fragmentation processes}
\label{sec3}

In the context of stabilized jellium model (SJM)\cite{pertran}
and self-consistent solution of the Kohn-Sham (KS)
equations\cite{KohnSham} in the density functional
theory\cite{Kohn64} with local spin-density approximation, we have
calculated the total energies\cite{PayamiJPC} of Ag clusters and
thereby obtained the energies needed for the fragmentation via
different decay channels. In the SJM, the discrete ions are
replaced by a uniform positive charge background of density
$n=3/4\pi r_s^3$ in which $r_s$ is the bulk value of the
Wigner-Seitz (WS) radius of the valence electrons of the metal
and for Ag, $r_s=3.02\; bohrs$. The geometry chosen for the
positive background is spherical. We have studied the binary
decay processes of positively charged Ag$_N^+$, Ag$_N^{2+}$
clusters containing up to $N=100$ atoms in all possible channels.
We have considered the following possible decay processes for
singly ionized Ag clusters
\begin{equation}
{\rm Ag}_N^{1+}\to {\rm Ag}_{N-p}^{1+} + {\rm Ag}_p^0,
\;\;\;\;\;\;\;\;p=1,2,\cdots,N-1. \label{s3eq1}
\end{equation}
For doubly charged clusters, the decays can proceed via two
different processes. The first one is the evaporation process
\begin{equation}
{\rm Ag}_N^{2+}\to {\rm Ag}_{N-p}^{2+} + {\rm Ag}_p^0,
\;\;\;\;\;\;\;\;p=1,2,\cdots,N-3 \label{s3eq2}
\end{equation}
and the second one is fission into two charged products
\begin{equation}
{\rm Ag}_N^{2+}\to {\rm Ag}_{N-p}^{1+} + {\rm Ag}_p^{1+},
\;\;\;\;\;\;\;\;p=1,2,\cdots,[N/2]. \label{s3eq3}
\end{equation}

In evaporation processes, the negativity of the difference
between total energies before and after fragmentation
(dissociation energy),
\begin{equation}
D^Z(N,p)=(E_{N-p}^{Z} + E_p^{0}) - E_N^{Z}, \label{eqdiss}
\end{equation}
is sufficient to have a spontaneous decay. However, in fission
processes, because of the possible existence of a fission barrier,
the negativity of the dissociation energy is not a sufficient
condition for the fission of the parent cluster. In Fig.
\ref{fig9}, the fission of a $Z$-ply charged $N$-atom cluster
into two clusters of respective sizes $N_1,\;N_2=N-N_1$, and
respective charges $z_1,\;z_2=Z-z_1$ is schematically shown. For
the fission, we have used the two-spheres approximation. $Q_f$ is
the energy release, $B_c$ is the fusion barrier which is equal to
the maximum energy of the Coulomb interaction of two
positively-charged conducting spheres, taking their
polarizabilities into account. $B_f$ is the fission barrier
height which is defined as
\begin{equation}\label{eqfisbar}
  B_f=-Q_f+B_c.
\end{equation}

The Coulomb interaction energy of two charged metal spheres has
been numerically calculated using the classical method of image
charges\cite{Naher}. The calculations show that the maximum of
the interaction energy is achieved for separations $d_0\ge
R_1+R_2$. The most favored channel for the evaporation is defined
as the channel for which the dissociation energy is minimum.
Similarly, the most favored channel for the fission has the
minimum of the barrier height.

\subsection{Calculation of the total energy of a cluster}
In the context of the SJM, the total energy of an $N$-atom
$Z$-ply charged cluster is given by

\begin{eqnarray}
E_{\rm SJM}\left[\nup,\ndo,n_+\right]&=& E_{\rm
JM}\left[\nup,\ndo,n_+\right]+\left(\varepsilon_M(r_s^B)+\bar
w_R(r_s^B,r_c^B)\right)\int d\rr\;n_+(\rr) \nonumber \\
  &&+\langle\delta v\rangle_{\rm WS}(r_s^B,r_c^B)\int
d\rr\;\Theta(\rr)\left[n(\rr)-n_+( \rr)\right], \label{s3eq4}
\end{eqnarray}
where
\begin{eqnarray}
E_{\rm
JM}\left[\nup,\ndo,n_+\right]&=&T_s\left[\nup,\ndo\right]+E_{xc}\left[\nup,\ndo
\right] \nonumber\\ &&+\frac{1}{2}\int
d\rr\;\phi\left([n,n_+];\rr\right)\left[n(\rr)-n_+(\rr)\right]
\label{s3eq5}
\end{eqnarray}
and
\begin{equation}
\phi\left([n,n_+];\rr\right)=\int
d\rrp\;\frac{\left[n(\rrp)-n_+(\rrp)\right]}{\left|\rr-\rrp\right|}.
\label{s3eq6}
\end{equation}
Here, $n=n_\uparrow+n_\downarrow$ which satisfies $\int d\rr
n(\rr)=N-Z$, and $n_+$ is the jellium density which satisfies
$\int d\rr n_+(\rr)=N$. $\Theta(\rr)$ takes the value of unity
inside the jellium background and zero, outside. The first and
second terms in the right hand side of Eq.(\ref{s3eq5}) are the
non-interacting kinetic energy and the exchange-correlation
energy, and the last term is the Coulomb interaction energy of
the system. The quantity $\langle\delta v\rangle_{\rm WS}$ is the
average of the difference potential over the Wigner-Seitz cell
and the difference potential, $\delta v$, is defined as the
difference between the pseudo-potential of a lattice of ions and
the electrostatic potential of the jellium positive background.
All equations throughout this paper are expressed in atomic units
($\hbar=e^2=m=1$), otherwise, they will be explicitly specified.
Using the Eq. (21) of Ref. [\ref{pertran}], this average value is
given by

\begin{equation}
\langle\delta v\rangle_{\rm
WS}(r_s^B,r_c^B)=\frac{3(r_c^B)^2}{2(r_s^B)^3}-
\frac{3}{10r_s^B}, \label{s3eq11}
\end{equation}
where, $r_c^B$ is the core radius of the pseudo-potential and is
given by\cite{PayamiJPC}

\begin{equation}
r_c^B=\frac{1}{3}(r_s^B)^{3/2}\left\{\left[
-2t_s(r_s)-\varepsilon_x(r_s)+ r_s \frac{\partial}{\partial
 r_s}\varepsilon_c(r_s)-\varepsilon_M(
 r_s)\right]_{r_s=r_s^B}\right\}^{1/2}.
\label{s3eq12}
\end{equation}
The quantities $t_s$, $\varepsilon_x$, and $\varepsilon_c$ are the
kinetic, exchange, and correlation energies per electron,
respectively\cite{Payami98}. The effective potential, used in the
self-consistent KS equations, is obtained by taking the
variational derivative of the SJM energy functional with respect
to the spin densities as

\begin{eqnarray}
v_{eff}^\sigma\left(\left[n_\uparrow,n_\downarrow,n_+\right];\rr\right)&=&
\phi\left(\left[n,n_+\right];\rr\right)+
v_{xc}^\sigma\left(\left[n_\uparrow,n_\downarrow\right];\rr\right)
 +\Theta(\rr)\langle\delta v\rangle_{\rm WS} (r_s^B,r_c^B),
\label{s3eq7}
\end{eqnarray}
where $\sigma=\uparrow,\downarrow$. By solving the KS equations
\begin{equation}
\left(-\frac{1}{2}\nabla^2+v_{eff}^\sigma(\rr)\right)\psi_i^\sigma(
\rr)=\varepsilon_i^\sigma
 \psi_i^\sigma(\rr),\;\;\;\;\;\;\;\sigma=\uparrow,\downarrow,
\label{s3eq8}
\end{equation}
and finding the self-consistent values for $\varepsilon_i^\sigma$
and $\psi_i^\sigma$, with

\begin{equation}
n(\rr)=\sum_{i(occ),\sigma=\uparrow,\downarrow}|\psi_i^\sigma(\rr)|^2,
\label{s3eq9}
\end{equation}
the total energy of a cluster is calculated.

\subsection{Results of Calculations}
After the self-consistent calculations, we have calculated the
total energies of Ag$_N^Z$ ($Z$=0,1,2) for different cluster
sizes $(1\le N\le 100)$. Then, using Eqs. (\ref{eqdiss}) and
(\ref{eqfisbar}), we have calculated the dissociation energies
and the fission barriers.

In Fig. \ref{fig10}, we have plotted the dissociation energies of
the most favored evaporation channels of the singly ionized
clusters. We have shown the most favored value of $p$ by $p^*$.
The solid small square symbols show the most favored values $p^*$
on the right vertical axis whereas, the corresponding
dissociation energies, $D^{1+}(N,p^*)$, are shown on the left
vertical axis by large open squares. As is seen in the figure,
there exist some maxima and minima. The maxima of the
$D^Z(N,p^*)$ correspond to the closed-shell Ag$_N^+$ clusters
with $N$=3, 9, 19, 35, 59,$\ldots$. These clusters have high
stabilities compared to their neighboring sizes. On the other
hand, the minima correspond to the sizes which decay into two
closed-shell clusters (for example, Ag$_{11}^+\rightarrow$Ag$_9^+
+$Ag$_2$). A negative value for the dissociation energy implies
that the cluster is unstable against the spontaneous decay.

In Fig. \ref{fig11}(a), we have shown the most favored products
Ag$_{p^*}^0$ and the dissociation energies $D^{2+}(N,p^*)$ for the
decay of {\rm Ag}$_N^{2+}$ via evaporation channel. It is seen
that the most favored products are mainly monomers, dimers and
octamers. On the other hand, here, the mean dissociation energy
is higher than that in the evaporation of singly ionized
clusters. That is, here, the number of clusters stable against
the spontaneous evaporation is larger than that in the singly
ionized case. Evaporation is not the only decay mechanism for
multiply charged clusters, and they can also decay via fission
processes in which both fragments are charged.

Figure \ref{fig11}(b) shows the barrier heights $B_f(N,p^*)$ for
 the most  favored fission channels
 ${\rm Ag}_N^{2+}\to {\rm Ag}_{N-p^*}^{+} + {\rm Ag}_{p^*}^{+}$. By
definition, the most favored fission channel has a minimum value
for the barrier height. As is seen, for small clusters, the
majority have negative barrier heights. That is, most of them are
unstable against spontaneous fission. However, as $N$ increases,
the number of clusters with negative barrier heights decreases
and beyond a certain size range, all the barrier heights become
positive.

A doubly charged cluster decays both via evaporation and fission.
At small sizes ($N<21$), the fission process dominates because,
the barrier heights for the fission are lower than the
dissociation energies for the evaporation. This situation is
shown in Fig. \ref{fig11}(c). As is seen, the competition between
the evaporation and fission starts at $N$=21. This competition
continues with some fluctuations until the evaporation dominates
completely. Since, the cluster sizes in the a ion-exchanged sample
are\cite{Gonella} of order $\sim$5 nm, which corresponds to $N\gg
21$, the fragmentations mostly proceed via the evaporation
mechanism.

\section{Conclusion}
\label{sec4}

We have studied the interaction of a CW laser beam with the
clusters embedded in glass matrix. The study consists of two
regimes of fast and slow. In both regimes, neutral clusters are
generated (creation of a 410 nm band peak). Depending on the
regime, different results are obtained. In the slow regime, the
absorption peak in the UV region of the spectrum is blue-shifted
relative to that of non-irradiated samples, which implies the
generation of smaller clusters. On the other hand, under the fast
regime, in addition to the fragmentations, clusters of $\mu$m
size are created. The experimental results show that heat
treatment of the ion-exchanged sample under the normal atmosphere
conditions also gives rise to the appearance of neutral clusters.
The theoretical calculations show that doubly ionized clusters of
size $N\gg 21$ decay mostly via the evaporation mechanism which
is one of the possible mechanism for the generation of neutral
clusters.

\begin{center}
{\large\bf Acknowledgement}
\end{center}
The authors are grateful to M. T. Tavassoli for his advice in some
experiments. They also appreciate A. Ghoreyshi for determination
of components of glass samples. The financial support of this
work is provided by research council of the Institute for
Advanced Studies in Basic Sciences.

\newpage

\begin{figure}
\caption{Schematic drawing of the ion-exchange process in sample
preparation. As a diffusion process, the Ag$^+$ ions in the molten
salt replace the Na$^+$ ions in the glass matrix. } \label{fig1}
\end{figure}

\begin{figure}
\caption{Experimental setup for irradiating the samples. The
circular interference pattern, shown on the top-left, is appeared
beyond the threshold power ($P_t\sim 4\;W$) } \label{fig2}
\end{figure}

\begin{figure}
\caption{Absorption spectra of the samples 1)- before, 2)- after
the irradiation by the high-power Ar$^+$ laser beam. The quantity
$D$ in the vertical axis corresponds to the dimension-less
optical density. The 410 nm absorption band peak is appeared
after the irradiation.}\label{fig3}
\end{figure}

\begin{figure}
\caption{Comparison of the absorption spectra under the fast
regime (curve 1) and the slow regime (curve 2). The spectra are
measured for the central parts of the interaction regions.
}\label{fig4}
\end{figure}

\begin{figure}
\caption{Absorption spectra from different parts of the
interaction region. The labels 1-12 correspond to the sub-regions
from the outer parts to the center of the circular interaction
region, respectively. The sub-regions are separated by
$0.2\;mm$.}\label{fig5}
\end{figure}

\begin{figure}
\caption{Plot of the normalized height of the 410 nm absorption
band peak as a function of the radial distance in the circular
interaction region. }\label{fig6}
\end{figure}

\begin{figure}
\caption{Absorption spectra of a sample. 1)- before and 2)- after
heat treatment at 350$\circ^C$ for 2 hours. After the heat
treatment, a weak 410 nm absorption peak is appeared and the UV
peak is red-shifted as well as increased}.
 \label{fig7}
\end{figure}

\begin{figure}
\caption{Micro-graphs of the interaction area of the sample after
irradiation under the fast regime. a)- The ring-shape area on the
front side of the sample with an island-like texture (50X). b)-
Area A in (a) is magnified (700X). c)- The rectangular region in
(b) is magnified (2000X). d)- Magnification of the area B in (a)
which belongs to the back side of the sample (300X), indicating
the formation of a dot-like texture. This photo is taken by the
dark-field method. e)- Magnified (700X) rectangular region of
photo (d). }\label{fig8}
\end{figure}

\begin{figure}
\caption{Fission barrier in the two-spheres approximation. The
parent $N$-atom $Z$-ply charged cluster decays into clusters of
sizes $N_1$ and $N-N_1$, with charges $Z_1$ and $Z_2$,
respectively. } \label{fig9}
\end{figure}

\begin{figure}
\caption{Dissociation energies, in electron volts, of the most
favored decay channels of singly ionized clusters are shown with
respect to the left vertical axis. The right vertical axis shows
the sizes of the fragments in the most favored channels.}
\label{fig10}
\end{figure}

\begin{figure}
\caption{a)- The same quantities as in Fig.\ref{fig10} for doubly
ionized clusters, b)- the fission barrier height in
electron-volts for the most favored fission channel as a function
of the cluster size. The right vertical axis shows the most
favored product sizes of the neutral clusters, c)- comparison of
the decay energies of via evaporation and fission mechanisms.
Competition starts at $N=21$ } \label{fig11}
\end{figure}

\newpage

\begin{table}
\begin{center}
\begin{tabular}{|c|c|c|c|c|c|c|c|c|c|}\hline
  Component & SiO$_2$ & CaO & Na$_2$O & MgO & Al$_2$O$_3$ & K$_2$O & S & Fe$_2$O$_3$ & P$_2$O$_5$
  \\ \hline
  Weight Percent(\%) & 80 & 9.41 & 4 & 3.3 & 2.2 & 0.41 & 0.2 & 0.11 & 0.11 \\ \hline
\end{tabular}
\caption{Components of the glass, in weight percent, which was
used in our experiments to produce the ion-exchanged samples.}
\label{tab1}
\end{center}
\end{table}

\begin{table}
\begin{center}
\begin{tabular}{|c|c|c|c|c|c|}\hline
  Wavelength(nm) & 457.9 & 476.5 & 488 & 496.5 & 514.5 \\ \hline
  Transmittance (\%) & 6.6 & 10.3 & 13.8 & 16.7 & 26.1 \\ \hline
\end{tabular}
\caption{Transmittance, in percent, of the ion-exchanged samples
for different lines in the pump beam.}
\label{tab2}
\end{center}
\end{table}

\end{document}